\documentclass[letterpaper, 10 pt, conference]{IEEEconf}  
\IEEEoverridecommandlockouts                              

	\usepackage{cite}
	\usepackage{amssymb,amsfonts,mathptmx}
	 
	\usepackage{amsthm}
	\usepackage{graphicx}
	\usepackage{enumerate}
	\usepackage{epsfig}
	\usepackage[font=small,labelfont=bf]{caption}
	\usepackage{amsmath} 
	\usepackage{bm}
	\usepackage{epstopdf,array}
	\usepackage{algorithm}
	\usepackage{algpseudocode}
	\usepackage{multirow,multicol}
	\let\labelindent\relax
	\usepackage{subfig,braket,enumitem,adjustbox,booktabs}
	\usepackage{color}
	\usepackage{float}
	\newfloat{algorithm}{t}{lop}
	\usepackage{pgf,tikz}
	\usetikzlibrary{arrows}

\newcommand{\real}{\mathbb{R}}

\newcommand{\T}{^{\top}}

\newcommand{\grad}{\nabla}
\newcommand{\tr}{\operatorname{tr}}

\newcommand{\Delxi}{\Delta \xi}
\newcommand\pd{\partial}

\newcommand{\kb}{\overline{k}}
\newcommand{\td}{\text{d}}
\newcommand{\Dc}{\mathcal{D}}

\newcommand{\longthmtitle}[1]{\mbox{}{\textit{(#1).}}}

\newcommand{\oprocendsymbol}{\hbox{$\bullet$}}
\newcommand{\oprocend}{\relax\ifmmode\else\unskip\hfill\fi\oprocendsymbol}

	\newtheorem{theorem}{Theorem}[section]
	\newtheorem{proposition}[theorem]{Proposition}
	\newtheorem{lemma}[theorem]{Lemma}
	\newtheorem{corollary}[theorem]{Corollary}
	\newtheorem{remark}[theorem]{Remark}

\usepackage{enumitem}

\begin{document}

\title{Control Co-design of systems with parabolic partial differential equation dynamics }
\author{Antika Yadav$^*$ and Prasad Vilas Chanekar
\thanks{$^{*}$Corresponding author.} 
\thanks{This research is supported by SERB Start-up Research Grant - SRG/2023/001636.}
\thanks{AY and PVC are with the Department of Electronics and
		Communication Engineering, Indraprastha Institute of Information Technology, New Delhi, India 110020, {\tt \{antikai,\,prasad\}}@iiitd.ac.in}}

\maketitle
\begin{abstract}
In this paper we study the control co-design (CCD) synthesis problem for a class of systems with parabolic partial differential equation (PDE) dynamics. We formulate CCD problem and finally derive an approximate CCD problem with matrix algebraic constraint. We then solve this approximate problem with gradient-based method and prove that the optimal solution also stabilizes the PDE system. We justify approach through numerical examples.  
\end{abstract}
\section{Introduction}
 Systems whose dynamics is governed by partial differential equations (PDEs) are relevant in various physical and engineering applications such as thermal regulation, structural vibration suppression, and chemical process control etc. A significant challenge in designing PDE systems is their infinite-dimensional nature. 
 
 
\textit{Literature Review:} Boundary state feedback control is a very effective method in control of PDE systems. In boundary feedback control the actuation is applied only at the boundary of the spatial domain. This makes their implementation very simple in practical applications. In this context, backstepping methods have been widely explored in the literature for PDE control \cite{boskovic2001boundary},\cite{smyshlyaev2005backstepping}. In \cite{Li2017FTStability}  the problem of finite-time (FT) stability and stabilization for distributed parameter systems is addressed.  In \cite{9980532}  a reinforcement learning-based boundary control strategy which  uses solution to a spatial Riccati-like equation for parabolic PDEs is presented. In \cite{9134871} a saturated feedback control has been derived, which locally stabilizes a linear reaction-diffusion equation. \cite{8392433} addressed distributed actuator selection by proposing a primal–dual algorithm that achieves optimality. 

 System design parameters and control parameters are inter-connected through the system dynamics. The traditional approach of first optimizing plant parameters and then synthesizing control gains may not lead to the best system. By optimizing system design and control gains simultaneously may lead to optimal system. This approach is known as Control Co-design (CCD) process. While CCD has been extensively studied for lumped-parameter systems (i.e., ODE-based models)\cite{10.1115/1.4027335},\cite{8812969}, its extension to PDE systems is largely unexplored. The existing literature regarding CCD is mainly concentrated on actuator/sensor location as design variable  \cite{7946139},\cite{15M1014759}. \cite{18M1171229}.  This research gap is the motivation of our work.

\textit{Statement of Contribution:} In this work, we propose the CCD problem for a class of systems with dynamics modeled by one dimensional parabolic partial differential equations. Using spatial discretizations we convert the PDE constrained, time-dependent problem to a time-dependent approximate optimization problem with ordinary differential equation (ODE) constraints. Deriving a PDE stability condition we then convert the time-dependent approximate problem CCD  problem into a problem with algebraic matrix constraints. We develop constraint gradient descent method to solve the approximate CCD method. Finally we prove the CCD solution computed for the approximate CCD problem is also a stabilizing feasible CCD solution to the original PDE system. Finally we justify our theory through numerical examples. 





\section{Preliminaries and Problem Statement}\label{prelim}
Consider\footnote{\textit{Notation:}$\mathbb{R}$ denotes the set of real numbers. A matrix $A \in \mathbb{R}^{m \times n}$ has $m$ rows and $n$ columns. $A \in \mathbb{R}^{m}$ represents m - dimensional vector. The symbols $A^\top$, $\operatorname{tr}(A)$, and $||A||$ represent the transpose, trace, and norm of the matrix $A$, respectively. For a symmetric matrix $P \in \mathbb{R}^{n \times n}$, $P \succeq 0$ (resp. $P \succ 0$) indicates that $P$ is positive semi-definite (resp. positive definite). $x_{\theta}$ denotes the partial derivative of x with respect to  $\theta$ i.e. $\frac{\partial x}{\partial \theta}$. The space \( L^2(0,1) \) is a Lebesgue space of square-integrable functions over the interval \( (0,1) \) defined as: $ L^2(0,1) := \left\{ f : (0,1) \to \mathbb{R} \;\middle|\; \int_0^1 |f(\xi)|^2 \, d\xi < \infty \right\}$. The Sobolev space is defined as $H^1(0,1) := \left\{ x \in L^2(0,1) \;\middle|\; x_\xi \in L^2(0,1) \right\}$. $C([0,\infty); H^1(0,1))$: space of continuous functions from $[0,\infty)$ into $H^1(0,1)$. We denote the PDE state $x(\xi,t)$ by $x$, $x _{\xi}(\xi,t)$ by $x_\xi$ and $x(\xi_0,t), \; x_\xi(\xi_0,t)$ by $x(\xi_0), \; x_\xi(\xi_0)$ respectively. For a function $f$, notation $\grad_{\theta}f$ denotes the gradient of $f$ with respect to $\theta$. J denotes the cost function. $\operatorname{diag}(\begin{bmatrix}a_1, a_2, \ldots, a_n\end{bmatrix})$ denotes the diagonal matrix with entries $a_1, a_2, \ldots, a_n$ on its main diagonal. Let $I$ denote the identity matrix of appropriate dimension.} the following system described by the parabolic PDE \cite{strauss2007partial} in one spatial dimension (1D),
\begin{align} \label{dynamics1}
    &x_{t}(\xi,t) = ax_{\xi\xi}(\xi,t) + b x(\xi,t),\;  x(\xi,0) = x_0(\xi) \nonumber\\
    &x_{\xi}(0,t) = u_1  , \;x_{\xi}(1,t) = u_2     \\
    &\text{where}\; \;  u_1 = k_1x(0,t),\; u_2 = k_2x(1,t)\nonumber
    \end{align}
where $ t > 0$ is the time variable, $\xi\in [0,\;1]$ is the spatial variable, $ a\in \real,\; b \in \real$ are the system parameters. 
 $x(\xi,t) \in C\big([0,\infty); H^1(0,1)\big)$ is the system state variable with $x_0(\xi)$ as the initial condition. $u = [u_1\; u_2]\T \in \real^2$ is the control vector applied to the boundary of the system. $u_1$ is applied at $\xi =0$ and  $u_1$ is applied at $\xi =1$. $k_1\in\real,\; k_2 \in \real$ are the stabilizing linear static feedback control gain. The system has the equilibrium state $x_e(\xi)\in \real.$ Note that $x_e = 0$ is one of the equilibrium points of the dynamics \eqref{dynamics1}.
 
 In general, the CCD optimization problem for system \eqref{dynamics1} is formulated as,
\begin{align}
 \label{ccd1}
&\notag \min_{d,\,k_1,\,k_2}  \; J = f_d(d)+ \int_{0}^{\infty} \int_{0}^{1}
\left(qx^2 + ru\T u\right)\td\xi \td t\\
 \notag \text{s.t.} \quad &x_t = ax_{\xi\xi} +bx,\; x(\xi,0) = x_0(\xi),\\
& x_\xi(0,t) = k_1x(0,t)=u_1,\; x_\xi(1,t) = k_2x(1,t)=u_2,   \\
&\notag x(\xi,t) \to x_e(\xi) \; \text{ as }\; t \to \infty,\\
&\notag (\xi,\;t) \in [0,\;1]\times [0,\; \infty), \; u = [u_1 \; u_2]\T.
\end{align}
Here, $d\in\real^{n_d}$ is the design parameter and $n_d$ is the number of design parameters, which in this case is two and $d = [a, b]^T$, $ f_d : \real^{n_d} \rightarrow \mathbb{R} $ is the design objective function.   $ q := q(\xi)\in \real,\;q(\xi) \geq 0 $, $ r := r(\xi) \in \real,\; r(\xi)> 0 $ are spatially distributed weighting functions. The term $\int_{0}^{\infty} \int_{0}^{1}
\left(qx^2 + ru\T u\right)\td\xi \td t$ quantifies the quadratic control cost of the system \eqref{dynamics1}. The constraint $x(\xi,t) \to x_e(\xi) \; \text{ as }\; t \to \infty$ ensures the stability of the co-designed system. The problem \eqref{ccd1} is a time-dependent, nonconvex and nonlinear optimization problem that is generally computationally challenging to solve. Note that for a fixed $d$, \eqref{ccd1} becomes an optimal control gain synthesis problem. Also, when $f_d(d) =0$ the design parameter $d$ influences the system CCD synthesis process through the dynamics \eqref{dynamics1}. 


\section{CCD Problem Reformulation}
\label{prob-reformulation}
In this section, we reformulate \eqref{ccd1} into an optimization problem with simple algebraic constraints. 

\subsection{Stability of the PDE system}
\label{pde-stability}
In this section, we derive stability conditions for system \eqref{dynamics1} by applying Lyapunov stability theory \cite{khalil2002nonlinear}.  The stability condition is in terms of the parameters $a,\:b, \: k_1,\:k_2$ and is presented in our next result.
\begin{theorem}\longthmtitle{Stability of PDE \eqref{dynamics1}}\label{stability1}
Consider the PDE \eqref{dynamics1} with the equilibrium state  $x_e=0$.
Then the system is asymptotically stable about  $x_e=0$ if \\
$\kb_1 - a < 0, \quad 5\kb_1 -a +4b < 0,\quad
2a k_2+\kb_1 + a <0 $\\
where $\kb_1 = \max\{0, \; -ak_1\}$.
\end{theorem}
\begin{proof}
Consider the Lyapunov function and its time-derivative
$$V = \frac{1}{2} \int_{0}^{1} x^2(\xi, t) \, \td\xi, \quad  V_t =\dot{V} = \int_{0}^{1} x x_{t}\td\xi $$
Now, using \eqref{dynamics1}, we get
\begin{align*}
\dot{V} & = a \int_0^1 x x_{\xi\xi} \, \td\xi + b \int_0^1 x^2 \, \td\xi \\
        &= a \left[ x(1)x_\xi(1) - x(0)x_\xi(0) \right] - a \int_0^1 x_\xi^2 \, \td\xi + b \int_0^1 x^2 \, \td\xi.
        \end{align*}
Using boundary conditions in \eqref{dynamics1},
\begin{align*}
 &x_\xi(0) = k_1 x(0) \; \text{and} \;x_\xi(1) = k_2 x(1),  \text{we get} \\
&\dot{V} 
= a k_2 x(1)^2 - a k_1 x(0)^2 - a \int_0^1 x_\xi^2 \, \td\xi + b \int_0^1 x^2 \, \td\xi \\
&\text{Using} \; \kb_1 = \max\{0, -ak_1\}, \text{we obtain}\\
&\dot{V} \leq a k_2 x(1)^2 +  \kb_1 x(0)^2 - a \int_0^1 x_\xi^2 \, \td\xi + b \int_0^1 x^2 \, \td\xi \\
&\text{Using Agmon's inequality:} \\ &x(0)^2 \leq x(1)^2 + 2 \sqrt{\int_0^1 x_\xi^2\,\td\xi} \sqrt{\int_0^1 x^2\,\td\xi}, \; \text{we get}\\
&\dot{V} \leq a k_2 x(1)^2 + \kb_1 \left[ x(1)^2 + 2 \sqrt{\int_0^1 x_\xi^2\,\td\xi} \sqrt{\int_0^1 x^2\,\td \xi} \right] \\
        &\quad - a \int_0^1 x_\xi^2 \, \td\xi + b \int_0^1 x^2 \, \td\xi \\
&\text{Using Young's inequality:}\; \text{if}\; a\geq 0,\; b\geq 0 \;\text{then}\\& 2ab \leq a^2 + b^2, \;\text{we get} \\
&\dot{V} \leq (a k_2 +  \kb_1) x(1)^2 +  \kb_1 \left[ \int_0^1 x_\xi^2 \, \td\xi + \int_0^1 x^2 \, \td\xi \right] \\
        &\quad - a \int_0^1 x_\xi^2 \, \td\xi + b \int_0^1 x^2 \, \td\xi \\
        &= (a k_2 +  \kb_1) x(1)^2 -  (-\kb_1 +a) \int_0^1 x_\xi^2 \, \td\xi \\
        &+ ( \kb_1 + b) \int_0^1 x^2 \, \td\xi \\
&\text{If}\;\;  \kb_1 - a < 0\;\; \text{and using Poincaré inequality: }\\&\text{for any x continuously differentiable on [0,\;1]} \\ &\int_0^1 x^2\,d\xi \leq 2x(1)^2 + 4\int_0^1 x_\xi^2\,\td\xi, \; \text{we get} \\
&\dot{V} \leq (a k_2 + \kb_1) x(1)^2 +  (\kb_1 + b ) \int_0^1 x^2 \, \td\xi \\
        &\quad + (a - \kb_1 ) \left[ \frac{x(1)^2}{2} -  \int_0^1 \frac{x^2}{4} \, \td\xi \right] \\
        &= \left(a k_2 + \frac{\kb_1}{2} + \frac{a}{2}\right) x(1)^2 + \left(\frac{5}{4}\kb_1 - \frac{a}{4} + b  \right) \int_0^1 x^2 \, \td\xi 
\end{align*}
If $\kb_1 -a < 0, \;2a k_2 + \kb_1 + a < 0, \; 5  \kb_1 - a + 4 b < 0$
 then $\dot{V} < 0$ 
implying asymptotic stability 
 \cite[Theorem 4.1]{khalil2002nonlinear}.
\end{proof}
For the homogeneous version of the PDE \eqref{dynamics1} the stability condition stated next.
\begin{corollary}\longthmtitle{Stability of homogeneous version of  \eqref{dynamics1}}\label{stability1-homogeneous}
 The homogeneous version of the system  \eqref{dynamics1} i.e., with $b=0$ is asymptotically stable about the equilibrium $x_e=0$ if 
 $$  \kb_1 - a < 0,\quad  5\kb_1 - a < 0,\quad
 2ak_2 + \kb_1 + a < 0$$
where $\kb_1 = \max\{0, -ak_1\}$.
\end{corollary}
Note that Theorem \ref{stability1} and Corollary \ref{stability1-homogeneous} are sufficient conditions only. Next, using the derived stability condition we reformulate the CCD optimization problem \eqref{ccd1}.
\subsection{Reformulated CCD problem}
\label{reform1}
We now reformulate \eqref{ccd1} as an optimization problem with ordinary differential equation (ODE) constraint. This is achieved by semi-discretizing the PDE in the spatial domain. To ensure stability, we  use Theorem \ref{stability1}. We now present our next result.
\begin{proposition}\longthmtitle{CCD problem discretization}\label{discrete1}
Consider the dynamics \eqref{dynamics1} and the problem \eqref{ccd1}. Let $\Delxi=\frac{1}{N-1}$ for some positive integer $N>1$. Then the system \eqref{dynamics1} can be rewritten as
$$\dot{X}(t) = A X(t), \quad X(0) = X_0,$$
for some $X(t) \in \mathbb{R}^N $, and $ A \in \mathbb{R}^{N \times N} $ with 
$$A = \scalebox{0.9}{$\frac{1}{\Delxi^2}$}\scalebox{0.65}{$
\begin{pmatrix}
-a -a k_1\Delxi + b\Delxi^2 & a & 0 & \cdots & 0 \\
a & -2a + b\Delxi^2 & a & \cdots & 0 \\
\vdots & \ddots & \ddots & \ddots & \vdots \\
0 & \cdots & a & -2a + b\Delta\xi^2 & a \\
0 & \cdots & 0 & a & -a + ak_2\Delta\xi + b\Delxi^2
\end{pmatrix}.$}$$ The cost $J$ in \eqref{ccd1} is rewritten as
$$J_d = f_d(d) + \int_0^\infty \left( X^\top Q_d X + U^\top R_d U \right) \td t$$ with some $U=KX,$
\begin{align*}
Q_d &= q\,\frac{\Delxi}{2} \,\operatorname{diag}\left(\begin{bmatrix}
   \frac{1}{2}& 1 \ldots 1& \frac{1}{2} 
\end{bmatrix}\right),\\
R_d &= r \,\operatorname{diag}\left(\begin{bmatrix}
   1&1
\end{bmatrix}\right),\;K  = \begin{pmatrix}
k_1 & 0 & \ldots & 0 \\
0 & \ldots &  0 & k_2
\end{pmatrix}.
\end{align*}
\end{proposition}
\begin{proof}
As $\xi \in [0, 1],$ partition [0, 1] in N-1 intervals $[\xi_{i-1}, \xi_i], \;i = 1, \dots,N$ of equal lengths with $\Delta \xi = \frac{1}{N-1}.$
Discretizing system~\eqref{dynamics1} in spatial domain\text{\cite{Thomas1995}}, where $x_i(t)$ represents $x(\xi_i, t)$ we have 
\begin{align}  \label{discretized system}  
 \dot{x_i}(t)  = a\frac{x_{i+1}(t)-2x_{i}(t)+x_{i-1}(t)}{(\Delta\xi)^2} + bx_i(t) \notag\\
 = a\frac{x_{i+1}(t)}{(\Delta\xi)^2}+ a\frac{x_{i-1}(t)}{(\Delta\xi)^2}
 +\left(b-\frac{2a}{(\Delta\xi)^2}\right)x_{i}(t)\\
 i = 1,2,....,N \notag
 \end{align}
 Now, to apply the Robin boundary conditions \eqref{dynamics1} using finite differences, we introduce \emph{ghost nodes} at \(i = 0\) and \(i = N+1\), located just outside the physical domain. These fictitious nodes are used to approximate spatial derivatives at the boundaries and are eliminated using the boundary conditions.
 Using backward and forward difference \cite{Thomas1995} at $\xi = 0 ,\; \xi = 1 $ respectively, we get
 \begin{align} \label{discretitized boundary condition}
\frac{x_1(t) - x_0(t)}{\Delta\xi} = k_1x_1(t),\;\;
 \frac{x_{N+1}(t) - x_N(t)}{\Delta\xi} = k_2x_N(t)
  \end{align}
 using \eqref{discretized system} and  \eqref{discretitized boundary condition} , we have
\begin{align} \label{system of odes}
\dot{x}_1(t) &=  a \dfrac{x_2(t)}{(\Delta\xi)^2} + \left( b + \dfrac{-a - a k_1 \Delta\xi}{(\Delta\xi)^2} \right) x_1(t),\notag\\ 
\dot{x}_i(t) &= a \dfrac{x_{i+1}(t) + x_{i-1}(t)}{(\Delta\xi)^2} + \left( b - \dfrac{2a}{(\Delta\xi)^2} \right) x_i(t);\\
& i = 2, \ldots, N-1,\notag\\ 
\dot{x}_N(t) &= a \dfrac{x_{N-1}(t)}{(\Delta\xi)^2} + \left( b + \dfrac{-a + a k_2 \Delta\xi}{(\Delta\xi)^2} \right) x_N(t) \notag, 
\end{align}
For $X = [ x_1 \; x_2 \ldots x_{N-1} \;x_N]\T \in \mathbb{R}^N$, \eqref{system of odes} becomes the system of ODEs as \\
$$\dot{X} = AX$$ 
where
$$
A = \frac{1}{\Delta\xi^2}
\scalebox{0.65}{$
\begin{pmatrix}
-a -ak_1\Delta\xi + b\Delta\xi^2 & a & 0 & 0 \\
a & -2a + b\Delta\xi^2 & a  & 0 \\
\vdots & \ddots & \ddots  & \vdots \\
0 & \cdots & a & a \\
0 & \cdots & 0 &  -a + ak_2\Delta\xi + b\Delta\xi^2
\end{pmatrix}
$}
$$
The cost function is 
$  J = f_d(d) + \int_{0}^{\infty} \int_{0}^{1}(qx^2+ru\T u) \td\xi \td t = f_d(d) + \int_{0}^{\infty} \int_{0}^{1}\left(qx^2+r(u^2_1 + u^2_2)\right) \td\xi \td t.$ Using trapezoidal rule \cite{atkinson1978introduction} and  $u_1 = k_1x(0), \; u_2 = k_2x(1),$  from \eqref{dynamics1} we have
\begin{align*}
 J_d = f_d(d) +& \int_{0}^{\infty}\left(\frac{\Delta\xi}{2}\left[ qx_1^2 + 2\sum_{i = 2}^{N-1}qx_i^2 + qx_N^2\right]\td t\right) \\ 
&+ (\int_{0}^{\infty}\left([rk_1^2x_1^2 + rk_2^2x_N^2]\td t \right)) 
\end{align*}
Now, for $U = [u_1 \;\; u_2]\T = [k_1x_1 \;\; k_2x_N]\T = KX$
$$ J_d =  f_d(d) +  \int_{0}^{\infty}\left( X\T (t)Q_dX(t) +U\T (t)R_dU(t)\right)\td t$$  $$= f_d(d) + \int_0^\infty X\T (Q_d+K\T R_dK)X \td t $$ 
where 
$$Q_d = q\frac{\Delta\xi}{2}  \operatorname{diag}\left(\begin{bmatrix}
   \frac{1}{2}& 1  \ldots  1& \frac{1}{2} 
\end{bmatrix}\right), \quad R_d = r \operatorname{diag}\left(\begin{bmatrix}
   1&1
\end{bmatrix}\right)$$

\end{proof}

\begin{remark}\longthmtitle{Relation to standard closed-loop system}\label{closed-loop}
		{\rm 
		In Proposition \ref{discrete1}, $A$ can be written as $A=A_0+BK$ where
$$ A_0 = \frac{1}{\Delta\xi^2}
\scalebox{0.65}{$
\begin{pmatrix}
-a  + b\Delta\xi^2 & a & 0 & \cdots & 0 \\
a & -2a + b\Delta\xi^2 & a & \cdots & 0 \\
\vdots & \ddots & \ddots & \ddots & \vdots \\
0 & \cdots & a & -2a + b\Delta\xi^2 & a \\
0 & \cdots & 0 & a & -a  + b\Delta\xi^2
\end{pmatrix}
$}, $$
$$B =  \frac{1}{\Delta\xi^2} \begin{pmatrix}
- a\Delta\xi & 0 \\
0 & 0 \\
\vdots & \vdots \\
0 &  0 \\
0 & a\Delta\xi
\end{pmatrix},\;
 K  = \begin{pmatrix}
k_1 & 0 & 0 & \cdots & 0 \\
0 & \cdots & 0 & 0 & k_2
\end{pmatrix}.
$$}		\oprocend
\end{remark} 
In Proposition \ref{discrete1} we observe that the matrix $A$ contains $K$ and $k_1,\;k_2$ are contained in $K$. Thus $A$ is a function of $d,\;,k_1,\;k_2$. Note that any reference to $K$ implies $K$ has the form as described in Remark \ref{closed-loop}.  Also, as $q\geq 0,\; r>0$ we have $Q_d\succeq 0$ and $R\succ 0$.  Using Proposition \ref{discrete1} and Theorem \ref{stability1} we can write the discrete, approximate version of the CCD problem \eqref{ccd1} as
\begin{align}
 \label{ccd2}
&\notag \min_{d,\,k_1,\,k_2}  \; J_d = f_d(d) + \int_0^\infty \left( X\T Q_d X + U\T R_d U \right) \td t\\
 \notag \text{s.t.} \quad &\dot{X} = A(d,k_1,k_2) X, \quad X(0) = X_0, \quad U=K(k_1,k_2)X,\\
& \kb_1 - a < 0, \quad 5\kb_1 -a +4b < 0,\\
&\notag 2a k_2+\kb_1 +a <0, \quad \kb_1 = \max\{0, \; -a k_1\},\\
&\notag \Vert X(t)\Vert \rightarrow 0 \quad \text{as} \quad t\rightarrow \infty.
\end{align}
The stability constraint $\Vert X(t)\Vert \rightarrow 0 \quad \text{as} \quad t\rightarrow \infty$ is achieved by ensuring that the optimized $A$ is Hurwitz \cite{khalil2002nonlinear}. The problem  \eqref{ccd2} is non-convex, nonlinear, time-dependent, and with the abstract stability constraint is challenging to solve. From optimal control theory \cite{lewis2012optimal} we have
$$\int_0^\infty \left( X\T Q_d X + U\T R_d U \right) \td t = \tr(PX_0), X_0 = x_0^Tx_0$$
where $P\succ 0$ is the solution to
\begin{align}
    \label{lyap1}
    A\T P + P\, A + Q_d + K\T R_d K = 0.
\end{align}
As $Q_d\succeq 0$ and $R_d\succ 0$, \eqref{lyap1} has a positive-definite solution when $A$ is Hurwitz \cite{lewis2012optimal}.  Using \eqref{lyap1}, the problem \eqref{ccd2} is equivalently rewritten as
\begin{align}
 \label{ccd3}
&\notag \min_{d,\,k_1,\,k_2}  \; J_f = f_d(d) + \tr(P\,X_0)\\
 \notag \text{s.t.} \quad &A\T P + P\, A + Q_d + K\T R_d K = 0, P \succ 0 \\
& \kb_1 - a < 0, \quad 5\kb_1 -a +4b < 0,\\
&\notag 2a k_2+\kb_1 +a <0, \quad \kb_1 = \max\{0, \; -a k_1\}.
\end{align}
The problem \eqref{ccd3} is time-independent, nonconvex, and  nonlinear with algebraic constraints. \eqref{ccd3}  is iteratively solved in a computationally tractable manner using a gradient-based method presented in the next section.
\section{Solution Procedure}
\label{sol-procedure1}
In this section we develop a gradient descent procedure \cite{Bertsekas}  to  compute an optimal solution to the problem \eqref{ccd3}. We first define the following set,
\begin{align}
      \label{set-def1}
         \Dc=\Set{(d,\,k_1,\,k_2)| \begin{array}{ll}
     (d,\,k_1,\,k_2)\in \real^{n_d} \times \real \times \real, \\
     \kb_1 - a < 0, \; 5\kb_1 -a +4b < 0,\\
2a k_2+\kb_1 +a <0,\\
\kb_1 = \max\{0, \; -ak_1\},\\
A(d,\,k_1,\,k_2)\text{  is Hurwitz}.
\end{array}}.
  \end{align}
  Note that for any $(d,\,k_1,\,k_2)\in\Dc$ implies the existence of a solution $P\succ 0$ to \eqref{lyap1} \cite{khalil2002nonlinear}.  Using \eqref{set-def1}, the problem \eqref{ccd3} is written as
  \begin{align}
 \label{ccd4}
& \min_{(d,\,k_1,\,k_2)\in \Dc}  \; J_f = f_d(d) + \tr(P\,X_0).
\end{align}
Note that \eqref{ccd4} is the `\textit{approximate}' version of the CCD problem \eqref{ccd1}. 
We develop a gradient-based method to solve \eqref{ccd4}. We first compute the gradient of $J_f$ with respect to $d,\;k_1,\;k_2$ in the next result.
\begin{lemma}\longthmtitle{Gradient computation}\label{gradient1} Consider the system \eqref{dynamics1}, CCD problem \eqref{ccd1} and its final approximation \eqref{ccd4} with $d\in\real^{n_d}$. Then the gradients $\frac{\pd J_f}{\pd d_j}=\frac{\pd f_d}{\pd d_j}+\tr\big(\frac{\pd P}{\pd d_j}\big)$ for $j=1,\dots,n_d$ and $\frac{\pd J_f}{\pd k_i}=\tr\big(\frac{\pd P}{\pd k_i}\big)$ for $i=1,2$ computed at some $(d^0,\,k_1^0,\,k_2^0)\in\Dc$. Where  $\frac{\pd P}{\pd d_j}$ is the solution to $$A\T \frac{\pd P}{\pd d_j}+\frac{\pd P}{\pd d_j}A+\frac{\pd A\T}{\pd d_j}P+P\frac{\pd A}{\pd d_j}=0,$$
and $\frac{\pd P}{\pd k_i}$ is the solution to $$A\T \frac{\pd P}{\pd k_i}+\frac{\pd P}{\pd k_i}A+\frac{\pd A\T}{\pd k_i}P+P\frac{\pd A}{\pd k_i}+\frac{\pd K\T}{\pd k_i}R_dK+KR_d\frac{\pd K}{\pd k_i}=0.$$
\end{lemma}
\begin{proof}
Differentiating $J_f$ and \eqref{lyap1} partially with respect to $d_j$ and $k_i$ gives the required result. 
\end{proof}
Using Lemma \ref{gradient1} we next outline the constrained gradient descent procedure \cite{Bertsekas} to solve \eqref{ccd4}.
\begin{algorithm}[H]
\caption{CCD Algorithm} 
\label{alg:GradientDescent}
\begin{algorithmic}[1]
\Require  $(d^0,\, k_1^0,\, k_2^0)\in \Dc$,  $Q_d$, $R_d$, $X_0$,  $\epsilon$, $\epsilon_1$, $\sigma$, $\beta$
\Ensure  $d^*$, $k_1^*$, $k_2^*$
\State Set: $j \gets 0$,  $k^j \gets [{d^j}\T,\;k_1^j,\; k_2^j]\T$
\State Using \eqref{lyap1} compute $J_f^j \gets \text{tr}(P_j X_0)$, 
\State Set: $J_f^{j-1} \gets -1$
\State Compute  $\nabla_k J_f^j$ using Lemma \ref{gradient1}
\While{ $\|\nabla_k J^j\| \geq \epsilon$ or $\vert J_f^j-J_f^{j-1}\vert \geq \epsilon_1$ }
    \State Compute $k^{j+1}$, $J_f^{j+1}$ from Algorithm \ref{alg:Armijo}
    \State Set: $k^j\gets k^{j+1}$, $J_f^{j-1}\gets J_f^j$, $J_f^j\gets J_f^{j+1}$
    \State $j\gets j+1$
    \State Compute  $\nabla_k J_f^j$ using Lemma \ref{gradient1}
\EndWhile
\State \Return  $d^* \gets d^j$, $k_1^* \gets k_1^j$, $k_2^* \gets k_2^j$
\end{algorithmic}
\end{algorithm}
Next, we state the procedure for computing $k^{j+1}$ and $J_f^{j+1}$ in Step 5 of Algorithm \ref{alg:GradientDescent}. The procedure is based on the Armijo step size selection rule \cite{Bertsekas}.
\begin{algorithm}[H]
	\caption{Computation of $k^{j+1},\;J_f^{j+1}$}\label{alg:Armijo}
	\begin{algorithmic}[1]
		\Require $\beta,\; \sigma,\; \nabla_k J_f^j,\; J_f^j,\;k^j$
		\Ensure $k^{j+1},\;J_f^{j+1}$
        \State Set $s^j \gets 1$
        \Repeat
            \State Compute $k^{j+1} = k^j - s^j \nabla_k J_f^j$
            \State Form $(d^{j+1},k_1^{j+1},k_2^{j+1})$ from $k^{j+1}$            \If{$(d^{j+1},k_1^{j+1},k_2^{j+1})\in \Dc$}
              \State Compute $J_f^{j+1}$ using \eqref{lyap1}
            \If{$J_f^{j+1} \leq J_f^j - \sigma s^j \Vert\nabla_k J_f^j\Vert^2$}
                \State Go to Step 15
            \Else
                \State $s^j \gets \beta s^j$
            \EndIf
        \Else
             \State $s^j \gets \beta s^j$
        \EndIf            
        \Until{Conditions satisfied}
    \State \Return $k^{j+1},\;J_f^{j+1}$
	\end{algorithmic}	
\end{algorithm}
We now prove that the optimal solution computed from Algorithm \ref{alg:GradientDescent} for the `\textit{approximate}' CCD problem \eqref{ccd4} is a stabilizing solution to problem \eqref{ccd1}. 
\begin{corollary} \longthmtitle{Relation between solutions of actual and approximate CCD problems}\label{relation1}
Consider system \eqref{dynamics1} with its CCD problem \eqref{ccd1} and the `\textit{approximate}' version \eqref{ccd4}. Let,  $(d^*,k_1^*,k_2^*)\in\Dc$ be the optimal solution of \eqref{ccd4} computed from Algorithm \ref{alg:GradientDescent}. Then $(d^*,k_1^*,k_2^*)\in\Dc$ is a feasible stabilizing solution of \eqref{ccd1}.
\end{corollary}
\begin{proof}
    Proved from $(d^*,k_1^*,k_2^*)\in\Dc$ satisfies Theorem \ref{stability1}.
\end{proof}
A detailed analytical study of the CCD Algorithm \ref{alg:GradientDescent} regarding its convergence and optimality properties is part of our future work. 

\section{Examples}
The dynamics described \eqref{dynamics1} is the mathematical representation of real-world practical systems like heat transfer process,  the gas diffusion process, etc., \cite{strauss2007partial}. We apply our proposed theory to design two version of  system \eqref{dynamics1}, namely `homogeneous system' $(b=0)$ and `nonhomogeneous system' $(b\neq 0)$. For each case we solve three types of design problems. 
\begin{enumerate}[leftmargin=0.5cm, labelsep=0.3cm, align=left]
    \item[\textbf{Case 1}:] Given $a,\,b$ compute optimal $k_1, \,k_2$.
    \item[\textbf{Case 2}:] Given $b$ compute optimal $a,\,k_1, \,k_2$ with $d=a$ and $f_d(a) = 0$.
    \item[\textbf{Case 3}:] Given $b$ compute optimal $a,\,k_1, \,k_2$ with $d=a$ and $f_d(a) = a^2$.
\end{enumerate}
We use Algorithm \ref{alg:GradientDescent} along with Algorithm \ref{alg:Armijo} for the CCD synthesis process.
The values of the necessary parameters required for the CCD computation process are as follows: 
\begin{align*}
    &q =1,\; r =10^4, \; \epsilon=10^{-3},\; \epsilon_1=10^{-6},
\sigma=0.3,\;\beta=0.3,\\ &X_0 = I, \;x_0(\xi) = J_0(n\xi), \;\text{which is defined in terms of  the} \\ &\text{bessel function \cite{watson1995treatise}}\; J_0(\xi) \;\;\text{and its first zero }\; n = 2.405.
\end{align*}
The initial control gain values are taken as $k_1^0 = 7 ,\; k_2^0 = -5$. For Case 2 and Case 3, the initial design parameter is taken as $a^0 = 10 $ For the `nonhomogeneous system' we take $b=-1$.  Our numerical analysis procedure is as follows.
\begin{itemize}
    \item We compute the initial and optimal values of the cost function in \eqref{ccd4}, i,e., $J_f^0,\;J_f^*$.
    \item We compute the  optimal variable values $a^*,\;k_1^*,\; k_2^*$.
    \item We solve the PDE  numerically using the MATLAB function `\textit{pdepe}' \cite{MATLAB1} for the spatial domain $\xi \in [0,\;1]$ and the time domain $t\in [0, \;200]$. The spatial domain was discretized in $25$ parts and the time domain was discretized in $500$ parts.
    \item We compute the initial and the final cost function values in \eqref{ccd1} i.e., $J^0,\;J^*$. This is done by numerically computing the double integral using the MATLAB function `\textit{trapz}'. The final cost is computed for $a^*,\;k_1^*,\; k_2^*$.
    \item For \textbf{Case 3}, we also compute the initial and the final control objective  value (denoted by $J_u^0,\; J_u^*$) in \eqref{ccd1} separately i.e., value of the double integral only. 
\end{itemize}
All computations are performed using MATLAB \cite{MATLAB1}. We now state our results.

\begin{figure}[h]
    \centering
    \includegraphics[height=7cm,width=9cm]{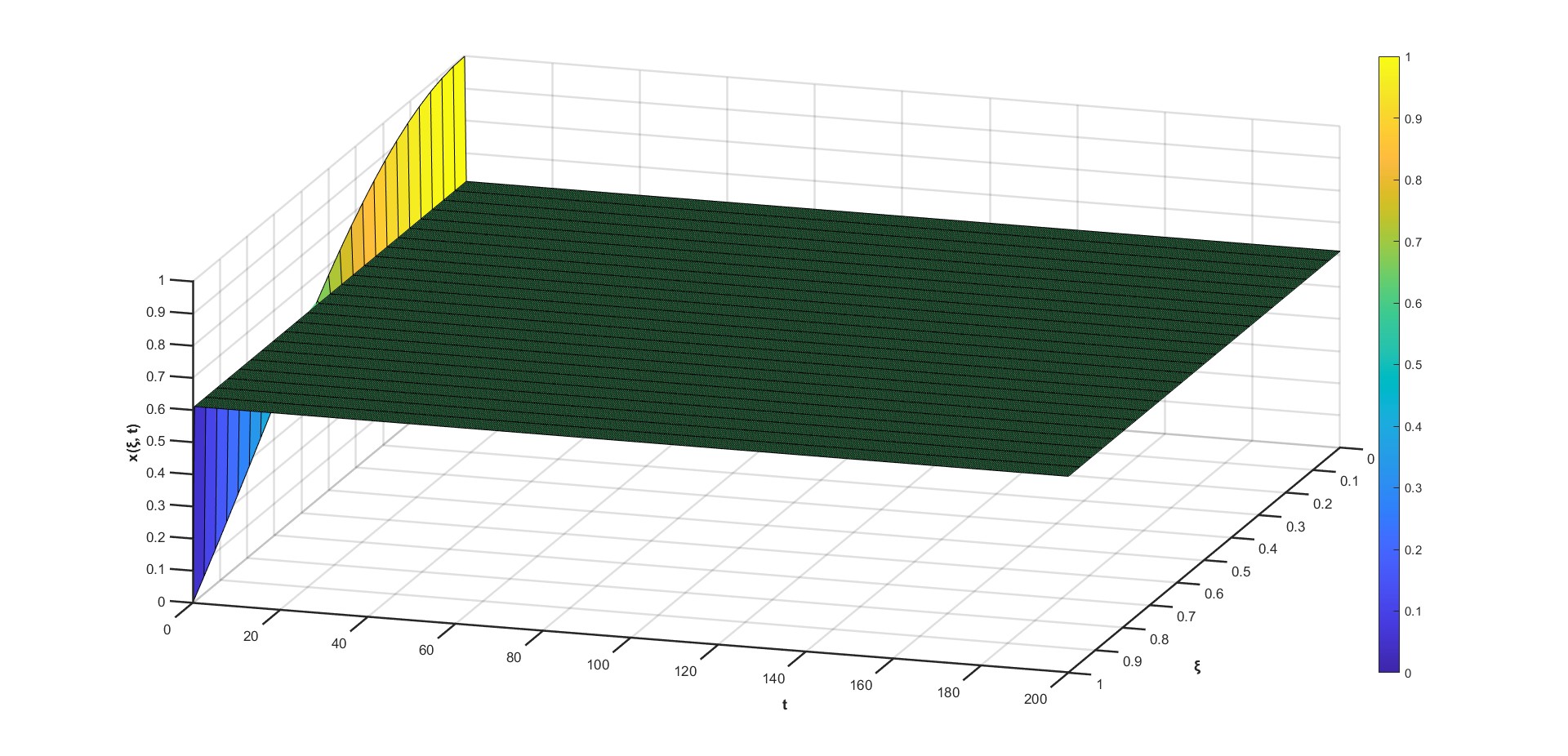}
    \caption{ $a = 10,\;b = 0,\;k_1 = 0,\; k_2 = 0$}
    \label{fig:6.1}
\end{figure}

{\textit{A. Homogeneous system $b=0$:}}
 The initial uncontrolled system i.e., $k_1=0,\,k_2=0$ is not asymptotically stable as shown in Fig. \ref{fig:6.1}.  The results are as follows,

\begin{enumerate}[leftmargin=0.5cm, labelsep=0.3cm, align=left]
   \item[\textbf{Case 1}:] $k_1^* = 2.12 , \,k_2^*= -0.50$, $J_f^0=276.6,\;J_f^*=53.77$, $J^0=98211,\;J^*=9961$.
   \item[\textbf{Case 2}:] $a^*=13.61,\;k_1^* =1.93 , \,k_2^*=-0.50$, $J_f^0=276.6,\;J_f^*=35.98$, $J^0=98211,\;J^*=8433$.
   \item[\textbf{Case 3}:] $a^*=9.75,\;k_1^* =2.02 , \,k_2^*=-0.50$, $J_f^0=376.6 ,\;J_f^*= 147.6$, $J^0= 118211,\;J^*=28142$, $J_u^0=98211,\;J_u^*= 9114$.
\end{enumerate}

\begin{figure}[h]
    \centering
    \includegraphics[height=7cm,width=9cm]{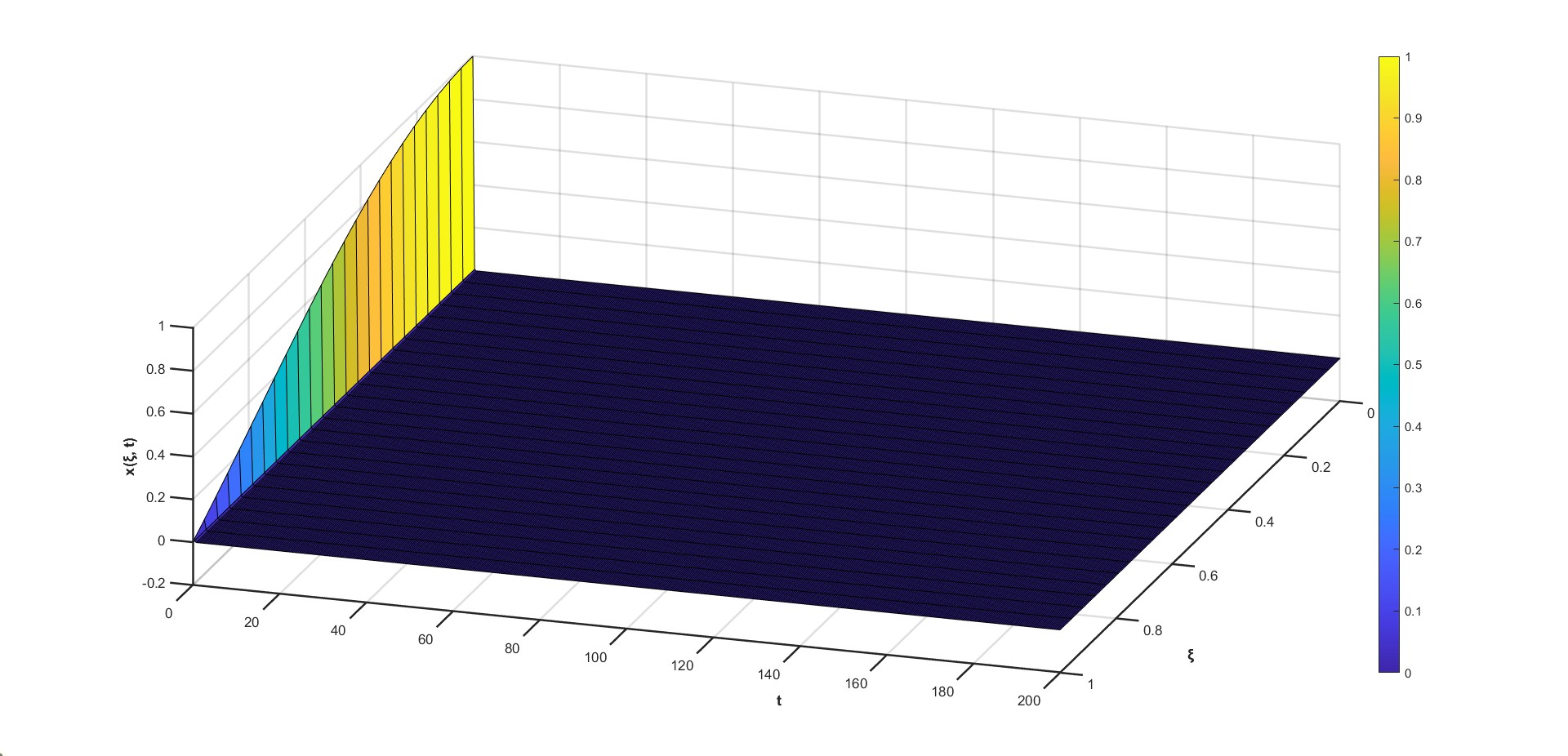}
    \caption{ $a = 10,\; b = 0,\;k_1 = 7,\; k_2 = -5$}
    \label{fig:6.2}
\end{figure}

As it can be seen form the Fig. \ref{fig:6.2} that the controlled system with $ k_1 = 7, k_2 = -5 $ makes the system asymptotically stable whereas the initial uncontrolled system with $k_1 = 0 , k_2 = 0 $ was not asymptotically stable, which is seen from Fig. \ref{fig:6.1}.

\begin{figure}[t]
    \centering
    \includegraphics[height=7cm,width=9cm]{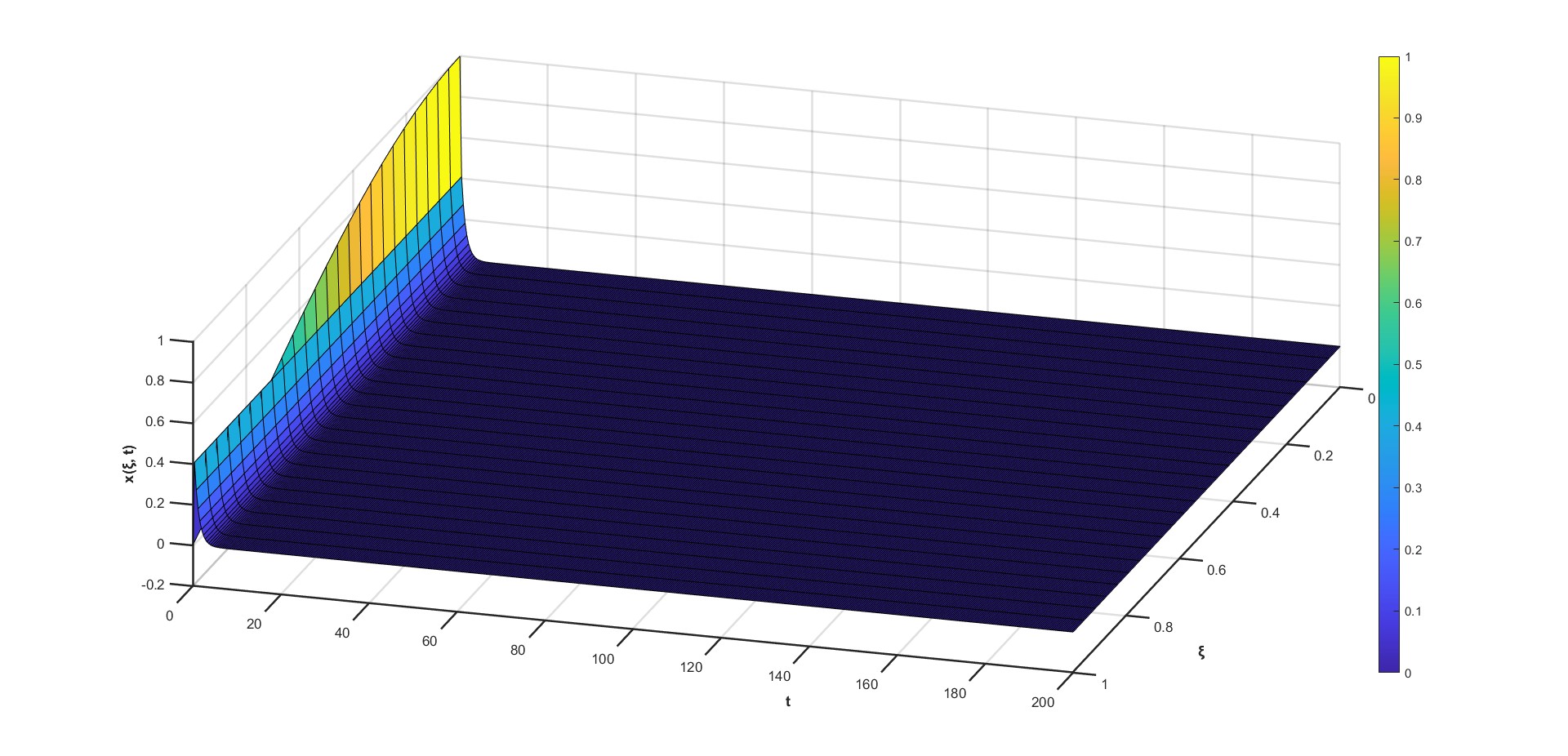}
    \caption{ $a = 10,\;b = -1,\;k_1 = 0,\; k_2 = 0$}
    \label{fig:6.3}
\end{figure}

{\textit{B. Nonhomogeneous system $b \neq
 0$:}}
The initial uncontrolled system i.e., $k_1=0,\,k_2=0$ is already asymptotically stable as shown in Fig. \ref{fig:6.3}.  The results are as follows,
\begin{enumerate}
\addtolength{\itemindent}{0.5cm}
   \item[\textbf{Case 1}:] $k_1^* = 2.13, \,k_2^*=-0.50$, $J_f^0=275,\;J_f^*=52.44$, $J^0=9823,\;J^*=9490$.
    \item[\textbf{Case 2}:] $a^*=13.57,\;k_1^* = 1.95, \,k_2^*=-0.50$, $J_f^0=275,\;J_f^*=35.58$, $J^0=98203,\;J^*=8001$.
    \item[\textbf{Case 3}:] $a^*=9.72,\;k_1^* = 2.04, \,k_2^*=-0.50$, $J_f^0=375,\;J_f^*=146.1$, $J^0=118203\;J^*=27630$, $J_u^0=98203,\;J_u^*=8735$.
\end{enumerate}

\begin{figure}[t]
    \centering
    \includegraphics[height=7cm,width=9cm]{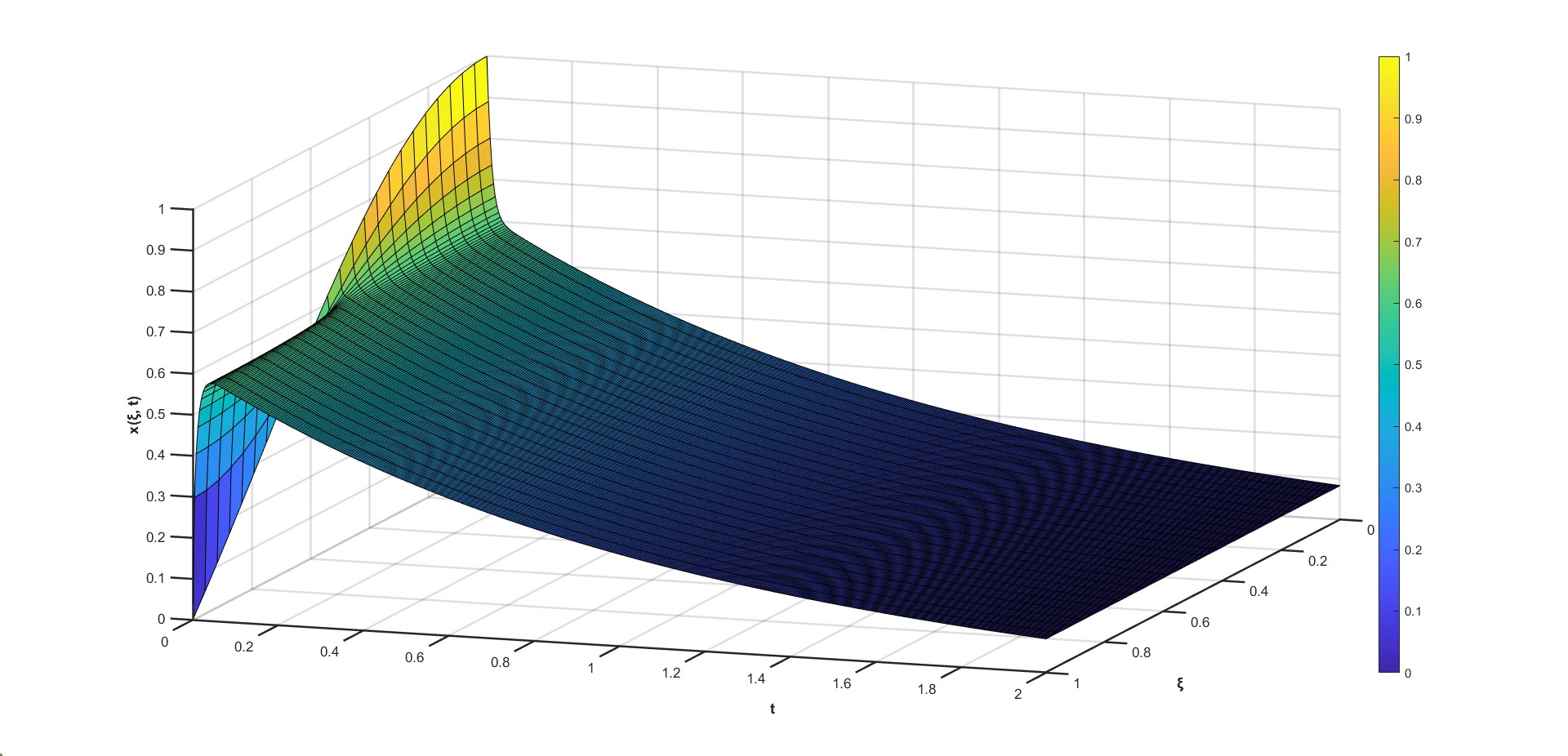}
    \caption{ $a = 10,\;b = -1,\;k_1 = 0,\; k_2 = 0$}
    \label{fig:6.5}
\end{figure}
\begin{figure}[t]
    \centering
    \includegraphics[height=7cm,width=9cm]{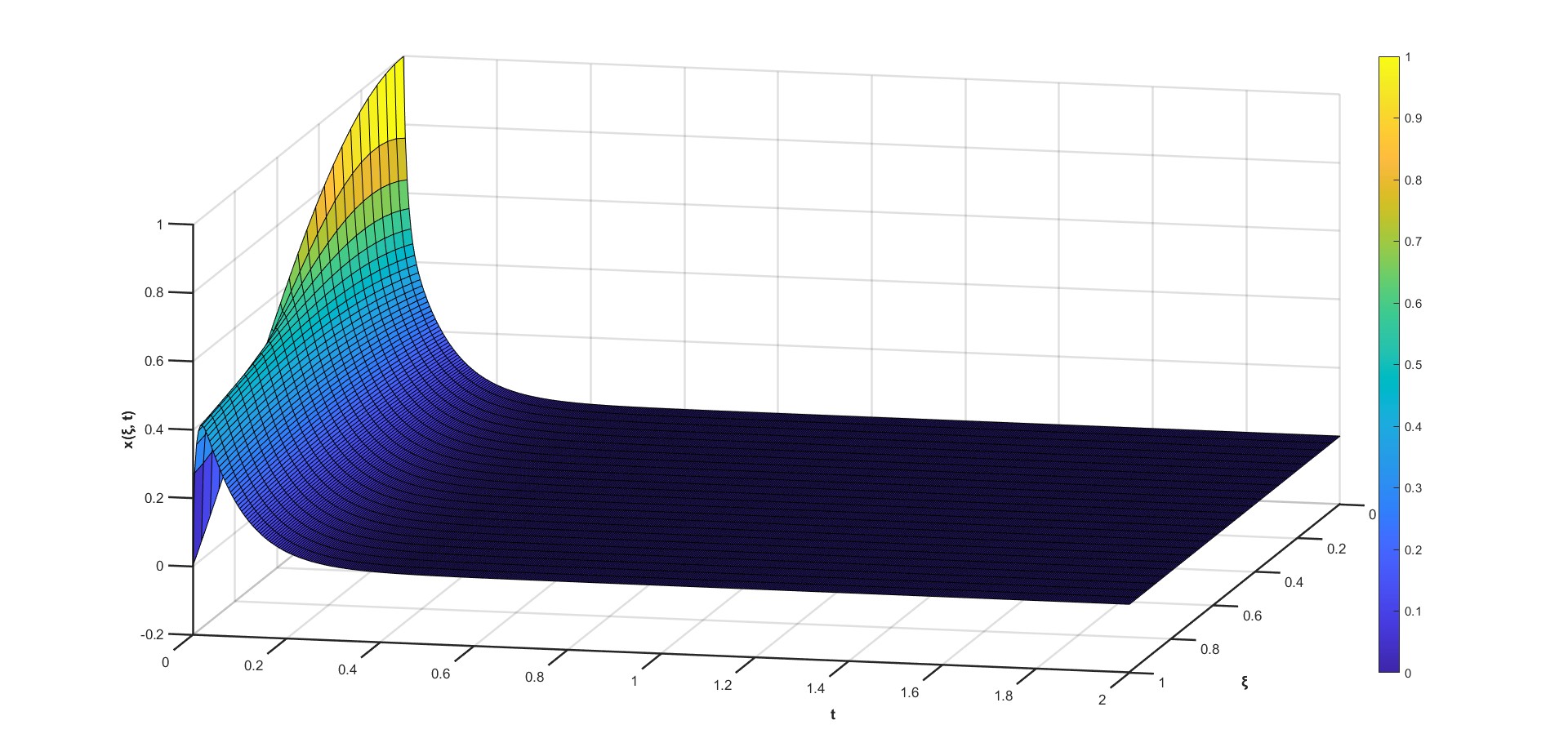}
    \caption{ $a = 10,\;b = -1,\;k_1 = 7,\; k_2 = -5$}
    \label{fig:6.6}
\end{figure}


 Fig.~\ref{fig:6.3} demonstrates that the uncontrolled system is already asymptotically stable in this case; however, applying the controller with \( k_1 = 7 \), \( k_2 = -5 \) significantly accelerates the convergence, as illustrated in Fig.~\ref{fig:6.5} and Fig.~\ref{fig:6.6}.

 For both homogeneous and nonhomogeneous systems, the control objective performance is better for CCD system  (Case 2 and Case 3) than only control design (Case 1). This shows efficacy of our proposed CCD theory.

\section{Conclusion}
In this paper we present a novel CCD system approach for class of systems with parabolic PDE dynamics. We derive novel stability condition for the PDE and later use it to develop a gradient-based CCD synthesis algorithm. We justify our approach through numerical examples. Future work includes study of analytical properties of convergence and optimality of the CCD algorithm.

\bibliographystyle{IEEEtran}
\bibliography{ref1}
\end{document}